# Weakly Supervised Deep Learning for COVID-19 Infection Detection and Classification from CT Images

Shaoping Hu [1]*, Yuan Gao [2,3]*, Zhangming Niu [3,4]*, Yinghui Jiang [4,5,6], Lao Li [4,5,6], Xianglu Xiao [3,5], Minhao Wang [4,5,6], Evandro Fei Fang[7], Wade Menpes-Smith[3], Jun Xia [8], Hui Ye [9†], and Guang Yang [10,11†]

[1] Radiology Department, Hospital of Wuhan Red Cross Society, Wuhan, China
[2] Institute of Biomedical Engineering, University of Oxford, UK
[3] Aladdin Healthcare Technologies Ltd, London, U.K
[4] Hangzhou Ocean's Smart Boya Co., Ltd
[5] Mind Rank Ltd, Hongkong, China
[6] IIAT
[7] Department of Clinical Molecular Biology, University of Oslo, Norway
[8] Radiology Department, Shenzhen Second People's Hospital, Shengzhen, China
[9] Department of PET-CT Center, Hunan Cancer Hospital, Changsha, China
[10] NHLI, Imperial College London, UK
[11] Royal Brompton Hospital, London, UK

* Co-first Authors.
† Corresponding Authors: Hui Ye and Guang Yang

## ABSTRACT

An outbreak of a novel coronavirus disease (i.e., COVID-19) has been recorded in Wuhan, China since late December 2019, which subsequently became pandemic around the world. Although COVID-19 is an acutely treated disease, it can also be fatal with a risk of fatality of 4.03% in China and the highest of 13.04% in Algeria and 12.67% Italy (as of 8th April 2020). The onset of serious illness may result in death as a consequence of substantial alveolar damage and progressive respiratory failure. Although laboratory testing, e.g., using reverse transcription polymerase chain reaction (RT-PCR), is the golden standard for clinical diagnosis, the tests may produce false negatives. Moreover, under the pandemic situation, shortage of RT-PCR testing resources may also delay the following clinical decision and treatment. Under such circumstances, chest CT imaging has become a valuable tool for both diagnosis and prognosis of COVID-19 patients. In this study, we propose a weakly supervised deep learning strategy for detecting and classifying COVID-19 infection from CT images. The proposed method can minimise the requirements of manual labelling of CT images but still be able to obtain accurate infection detection and distinguish COVID-19 from non-COVID-19 cases. Based on the promising results obtained qualitatively and quantitatively, we can envisage a wide deployment of our developed technique in large-scale clinical studies.

## INTRODUCTION

Coronavirus Disease 2019 (COVID-19) has been widespread worldwide since December 2019 [1][2]. It is highly contagious, and severe cases can lead to acute respiratory distress or multiple organ failure [3]. On 11 March 2020, the WHO has made the assessment that COVID-19 can be characterised as a pandemic. As of 8th April 2020, in total, 1,391,890 cases of COVID-19 have been recorded, and the death toll has reached 81,478 with a rapid increase of cases in Europe and North America.

The disease can be confirmed by using the reverse-transcription polymerase chain reaction (RT-PCR) test [4]. While being the gold standard for diagnosis, confirming COVID-19 patients using RT-PCR is time-consuming, and both high false-negative rates and low sensitivities may put hurdles for the presumptive patients to be identified and treated early [3][5][6].

As a non-invasive imaging technique, computed tomography (CT) can detect those characteristics, e.g., bilateral patchy shadows or ground glass opacity (GGO), manifested in the COVID-19 infected lung [7][8]. Hence CT may serve as an important tool for COVID-19 patients to be screened and diagnosed early. Despite its advantages, CT may share some common imagery characteristics between COVID-19 and other types of pneumonia, making the automated distinction difficult.

Recently, deep learning based artificial intelligence (AI) technology has demonstrated tremendous success in the field of medical data analysis due to its capacity of extracting rich features from multimodal clinical datasets [9]. Previously, deep learning was developed for diagnosing and distinguishing bacterial and viral pneumonia from thoracic imaging data [10]. In addition, attempts have been made to detect various chest CT imaging features [11]. In the current COVID-19 pandemic, deep learning based methods have been developed efficiently for the chest CT data analysis and classification [2][3][12]. Besides, deep learning algorithms have been proposed for COVID-19 monitoring [13], screening [14] and prediction of the hospital stay [15]. A full list of current AI applications for COVID-19 related research can be found elsewhere [16]. In this study, we will focus on the chest CT image based localisation for the infected areas and disease classification and diagnosis for the COVID-19 patients.

Although initial studies have demonstrated promising results by using chest CT for the diagnosis of COVID-19 and detection of the infected regions, most existing methods are based on commonly used supervised learning scheme. This requires a considerable amount of work on manual labelling of the data; however, at such an outbreak situation clinicians have very limited time to perform the tedious manual drawing, which may fail the implementation of such supervised deep learning methods. In this study, we propose a weakly supervised deep learning framework to detect COVID-19 infected regions fully automatically using chest CT data acquired from multiple centres and multiple scanners. Based on the detection results, we can also achieve the diagnosis for the COVID-19 patients. In addition, we also test the hypothesis that based on the CT radiological features, we can classify COVID-19 cases from community acquired pneumonia (CAP) and non-pneumonia (NP) scans using the deep neural networks we developed.

## MATERIALS AND METHODS

### Patients and Data

This retrospective study was approved by the institutional review board of the participating hospitals in accordance with local ethics procedures. Further consent was waived with approval. This study included 150 3D volumetric chest CT exams of COVID-19, CAP and NP patients, respectively. In total, 450 patient scans acquired from two participating hospitals between September 2016 and March 2020 were included for further analysis. All the COVID-19 patients were confirmed as positive by the RTPCR testing that were scanned from December 2019 to March 2020. CAP and other NP (no lung disease, lung nodules, chronic inflammation, chronic obstructive pulmonary disease) patients were randomly chosen from the participating hospitals between September 2016 and January 2020. CAP patients were laboratory confirmed bacterial culture positive cases or negative cases, e.g., with mycoplasma and viral pneumonia. NP patients were diagnosed with no lung disease or lung disease, e.g., lung nodules, chronic inflammation, chronic obstructive pulmonary disease and others.

COVID-19 patients were admitted from two hospitals in China, including 138 patients from Hospital of Wuhan Red Cross Society (WHRCH) and 12 patients from Shenzhen Second Hospital (SZSH). Both CAP and NP patients were recruited from SZSH. COVID-19 patients were obtained from either Siemens SIEMENS SOMATOM go.Now16 (WHRCH) or GE Revolution 256 (SZSH) CT systems. For the SIEMENS SOMATOM go.Now16 CT system, the scanning parameters were as follows: tube voltage = 130 kVp, automatic tube current modulation = 50 mAs, pitch = 1.5 mm, matrix = 512×512, slice thickness = 0.7 mm, field of view = 350 mm × 350 mm, and reconstructed slice thickness = 1 mm. For the GE Revolution 256 CT system, the scanning parameters were set as tube voltage = 120 kVp, automatic tube current modulation = 150 mAs, pitch = 1.375 mm, matrix = 512×512, slice thickness = 0.625 mm, field of view = 400 mm × 400 mm, and reconstructed slice thickness = 2 mm. All the CAP and NP patients were scanned using SIEMENS SOMATOM Emotion CT system with the main imaging parameters of tube voltage = 110 kVp, automatic tube current modulation = 70 mAs, pitch = 1.2 mm, matrix = 512×512, slice thickness = 1.2mm, field of view = 260 mm × 260 mm, and reconstructed slice thickness = 1.5 mm. Details are shown in Table 1.

Table 1: Imaging parameters of the CT systems used for COVID-19, CAP and NP patients.

| | COVID-19 Patients | COVID-19 Patients | CAP and non-pneumonia Patients |
|---|---|---|---|
| | WHRCH | SZSH | SZSH |
| Tube voltage | 130kV | 120KV | 110KV |
| Slice thickness | 0.7mm | 0.625mm | 1.2mm |
| Scanner | SIEMENS SOMATOM go.Now16 | GE revolution 256 | SIEMENS SOMATOM Emotion |
| Reconstructed slice thickness | 1 mm | 2mm | 1.5mm |
| Pitch | 1.5 | 1.375 | 1.2 |
| Matrix | 512×512 | 512×512 | 512×512 |
| Field of view | 350 mm × 350 mm | 400mm× 400 mm | 260mm× 260 mm |
| Automatic tube current modulation | 50 mAs | 150mAs | 70mAs |

**Dataset for Lung Segmentation**

In order to achieve a highly accurate lung segmentation that can facilitate the following infection detection and classification, we utilised an open dataset (TCIA dataset) [17] for training a deep neural network for the lung delineation. The data can be accessed from http://doi.org/10.7937/K9/TCIA.2017.3r3fvz08. In total, 60 3D CT lung scans were retrieved with manual delineations of the lung anatomy. These open datasets were made publicly accessible from the scans obtained by three different institutions: MD Anderson Cancer Centre, Memorial Sloan-Kettering Cancer Centre, and the MAASTRO clinic, with 20 cases from each institution. All the data were scanned with matrix = 512×512, the field of view = 500 mm × 500 mm, and reconstructed slice thickness varies at either 1 mm, 2.5 mm or 3 mm.

**Pre- and Post-Processing for Lung Segmentation**

Data pre-processing steps were performed to standardise data acquired from multiple centres and multiple scanners. Instead of normalising input slices into a pre-defined Hounsfield unit (HU) window, we designed a more flexible scheme based on previously proposed image enhancement methods [18][19]. Rather than clipping based on HU windows, we proposed to use a fixed-sized sliding window $W_{Q,S}$ (where $Q$ denotes the size of the window and $S$ denotes the step length of the sliding procedure) to find the range where covers most of the pixel values. This can reduce the bias of data acquired from different centres and different scanners. Loosely inspired by [20], we proposed a multi-view U-Net [21] based segmentation network for lung segmentation. Our multi-view U-Net based segmentation network consisted of a multi-window voting post-processing procedure and a sequential information attention module in order to utilise the information from each view of the 3D volume and reinforce the integrity of the 3D lung structure of the delineation results. Our lung segmentation model was trained, cross-validated and tested on the TCIA dataset with manual ground truth. The trained lung segmentation model was then used for inferencing the delineation of the lung anatomy of the COVID-19, CAP and NP patients included in this study.

### Detection and Classification Network

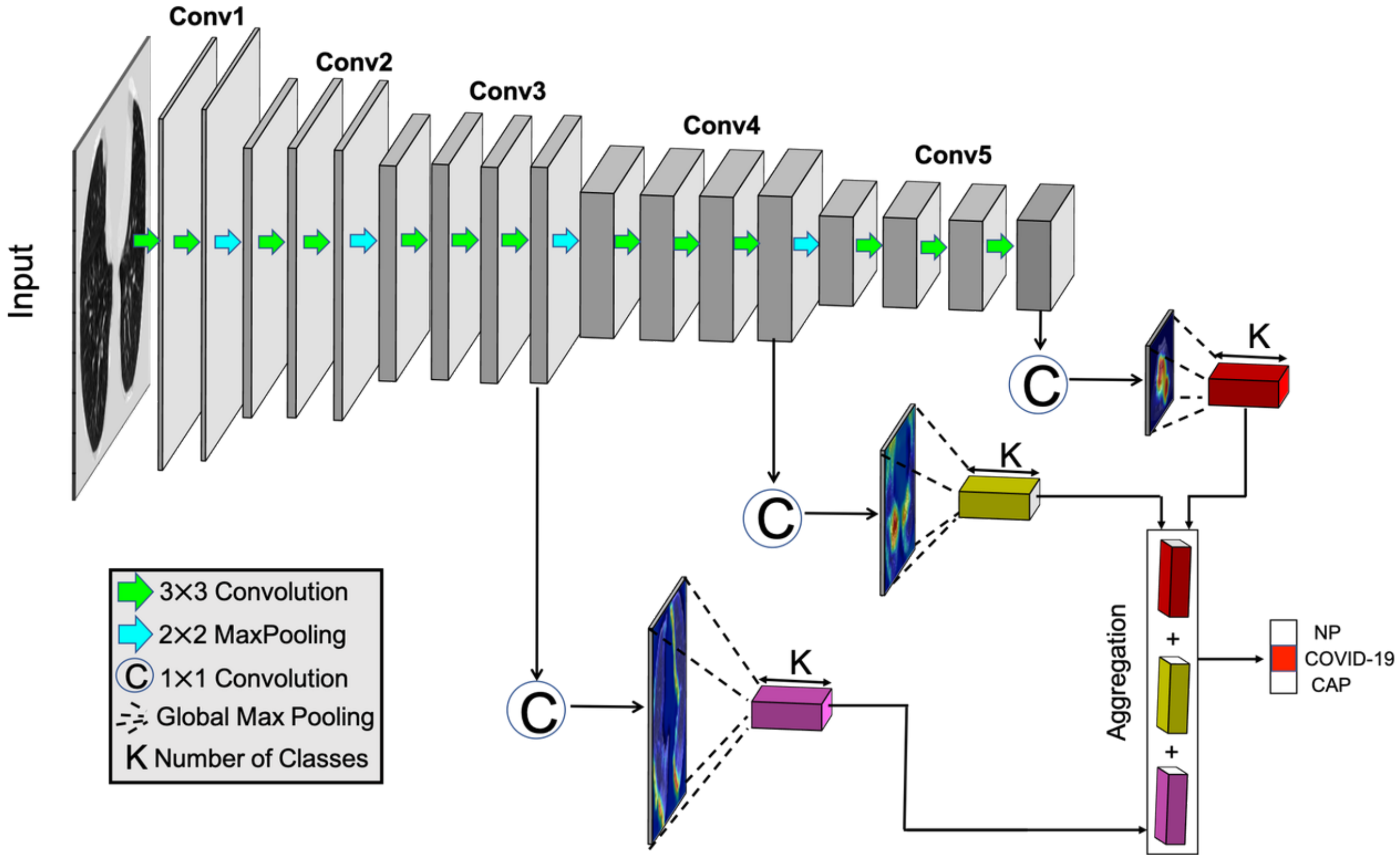

*Figure 1.: Network architecture of our proposed weakly supervised multi-scale learning framework for COVID-19/NP/CAP classification and lesions detection.*

Inspired by the VGG architecture [22], we adopted the configuration that increased CNN depth using small convolution filters stacked with non-linearity injected in between, as depicted in Figure 1. All convolution layers consisted of 3×3 kernels, batch normalisation and Rectified Linear Units. The proposed CNN was fully convolutional consisting of five convolutional blocks, i.e., Conv1, Conv2, Conv3, Conv4 and Conv5 in the backbone architecture. The full architecture, using shorthand notation, is 2× *C*(32,3,1)-*MP*-2× *C*(64,3,1)-*MP*-3× *C*(128,3,1)-*MP*-3× *C*(256,3,1)-*MP*-3× *C*(256,3,1)-*MP*, where *C(d,f,s)* indicates a convolution layer with *d* filters of spatial size *f*×*f*, applied to the input with stride *s*. *MP* represents non-overlapping max-pooling operation with a kernel size of 2×2.

### Multi-Scale Learning

From the previous findings using CT [23] [24] [25], it is known that infections of COVID-19 share the similar and common radiographic features as CAP, such as GGO and airspace consolidation. They frequently distribute bilaterally, peripherally in lower zone predominant, and the infectious areas can vary significantly in size depending on the condition of the patients. For example, in mild cases the abnormalities appear to be small, but in severe cases they appear scattered and spread around over a large area. Therefore, we proposed a multi-scale learning scheme to cope with variations of the size and location of the lesions. To implement this, we fed the intermediate CNN representations, i.e., feature maps, at Conv3, Conv4 and Conv5, respectively into the weakly supervised classification layers, in which 1×1 convolution was applied to mapping the feature maps down to the class score maps (i.e., class activation maps). We then applied a spatial aggregation with a Global Max Pooling (GMP) operation to obtain categorical scores. The scores vectors at Conv3, Conv4 and Conv5 level were aggregated by sum to make a final prediction with a Softmax function. We then trained the proposed model end-to-end by minimising the following objective function

$$L = -\frac{1}{N} \sum_{i=1}^{N} w_i f_i (S_c(x_i) - log \sum_{k=1}^{K} e^{S_k(x_i)}) \quad (1)$$

where there are $N$ training images $x_i$ and $K$ training classes. $S_k$ is the $k_{th}$ component in the score vector $\in \Re^K$, and $c$ is the true class of $x_i$. As we encountered an imbalanced classification, we added a class-balanced weighting factor $w_i$ to the cross-entropy loss, which was set by inverse class frequency, i.e., $w_i = \frac{1}{freq(c)}$. While this emphasised the importance of a rare class during training, it showed no difference between easy and hard examples. For instance, in mild COVID-19 slices, infectious or diseased regions are often very small and not prominent. Thus, they are prone to be misclassified as NP examples. To address this, we introduced another modulating factor, i.e., to down-weight easy examples and therefore focused the training on hard examples [26]: $f_i = (1 - P_c)^\gamma$, where $P_c$ is the true class posterior probability of $x_i$. Intuitively, the modulating factor can reduce the loss contribution from easy examples. This in turn increases the importance of correcting misclassified examples. When an example was misclassified and $P_c$ was small, the factor $f$ was near 1 and the loss was unaffected. As $P_c \to 1$, the factor went to 0 and the loss for well-classified examples was down-weighted. The parameter $\gamma$ is a positive integer which can smoothly adjust the rate at which easy examples are down-weighted. As $\gamma$ is increased the modulating effect of the factor $f$ is likely to be increased.

**Weakly Supervised Lesions Localisation**

After determining the class score maps and the image category in a forward pass through the network, the discriminative patterns corresponding to that category can then be localised in the image. A coarse localisation could already be achieved by directly relating each of the neurons in the class score maps to its receptive field in the original image. However, it is also possible to obtain pixel-wise maps containing information about the location of class-specific target structures at the resolution of the original input images. This can be achieved by calculating how much each pixel influences the activation of the neurons in the target score map. Such maps can be used to obtain a much more accurate localisation, like the examples shown in Figure 2.

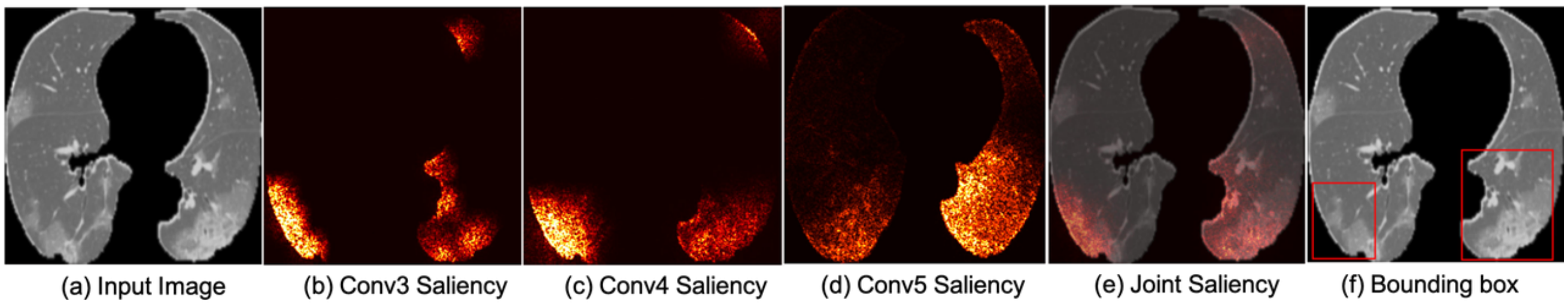

*Figure 2.: Examples of saliency maps for COVID-19 lesions localisation: (a) shows an example input image, (b) shows the saliency map obtained at Conv3, (c) shows the saliency map obtained at Conv4, (d) shows the saliency map obtained at Conv5, (e) shows the overlay of the joint saliency map (pixel-wise multiplication of the Conv3, Conv4 and Conv5 saliency maps) with the input image, and (f) shows the resulting bounding boxes.*

In the following, we will show how categorical-specific saliency maps can be obtained through the integrated gradients. Besides, we will also show how to post-process the saliency maps from which we can extract bounding boxes around the detected lesions.

### A. Category-Specific Saliency

Generally, suppose we have a flattened input image denoted as $x = (x_1, ..., x_n) \in \Re^n$ (number of pixels=n), category-specific saliency map can be obtained by calculating the gradient of the predicted class score $S(x)$ at the input $x$: $g = \frac{\partial S(x)}{\partial x} = (g_1, ..., g_n) \in \Re^n$, where $g_i$ represents the contribution of individual pixel $x_i$ to the prediction. In addition, the gradient can be estimated by back-propagating the final prediction score through each layer of the network. There are many state-of-the-art back-propagation approaches, including Guided-Backpropagation [27], DeepLift [28] and Layer-wise Relevance Propagation (LRP) [29]. However, Guided-Backpropagation method may break gradient sensitivity because it back-propagates through a ReLU node only if the ReLU is turned on at the input. In particular, the lack of sensitivity causes gradients to focus on irrelevant features and results in undesired saliency localisation. DeepLift and LRP methods tackle the sensitivity issue by computing discrete gradients instead of instantaneous gradients at the input. However, they fail to satisfy the implementation invariance because the chain rule does not hold for discrete gradients in general. In doing so, the back-propagated gradients are potentially sensitive to unimportant features of the models. To deal with these limitations, we employ a feature attribution method named "Integrated Gradients" [30] that assigns an importance score $\phi_i(S(x), x)$ (similar to pixel-wise gradients) to the $i_{th}$ pixel representing how much the pixel value adds or subtracts from the network output. A large positive score indicates that pixel strongly increases the prediction score $S(x)$, while an importance score closes to zero indicates that pixel does not influence $S(x)$. To compute the importance score, it needs to introduce a baseline input representing "absence" of the feature input, denoted as $x^{'} = (x^{'}_1, ..., x^{'}_n) \in \Re^n$, which in our study, was a null image (filled with zeros) with the same shape as input image $x$. We considered the straight-line path, i.e., point-to-point from the baseline $x^{'}$ to the input $x$, and computed the gradients at all points along the path. Integrated gradients can be defined as

$$\phi_i(S(x), x, x^{'}) = (x_i - x_i^{'}) \times \int_{\alpha=0}^{1} \frac{\partial S(x^{'} + \alpha(x - x^{'}))}{\partial x_i} d\alpha \qquad (2)$$

where $\alpha \in [0, 1]$. Intuitively, integrated gradients can obtain importance scores by accumulating gradients on images interpolated between the baseline value and the current input. The integral in Eq. 2 can be efficiently approximated via a summation of the gradients as:

$$\phi_i(S(x), x, x^{'}) \approx (x_i - x_i^{'}) \times \sum_{n=1}^{m} \frac{\partial S(x^{'} + \frac{n}{m} \times (x - x^{'}))}{\partial x_i} \times \frac{1}{m} \qquad (3)$$

where $m$ is the number of steps in the Riemann approximation of the integral. We compute the approximation in a loop over the set of inputs, i.e., for $n = 1, ..., m$. The integrated gradients are computed at different feature levels, in our experiments, which are Conv3, Conv4 and Conv5 respectively, as shown in Figure 2(b), Figure 2(c) and Figure 2(d). Then, a joint saliency can be obtained, as depicted in Figure 2(e), by pixel-wise multiplication between the multi-scale integrated gradients.

### B. Bounding Box Extraction

Next, we post-processed the joint saliency map from which a bounding box can be extracted. Firstly, we took the absolute value of the joint saliency map and blurred it with a $5 \times 5$ Gaussian kernel. Then, we thresholded the blurred saliency map using the Isodata thresholding method [31] that it iteratively decided a threshold segmenting the image into foreground and background, where the threshold was midway between the mean intensities of sampled foreground and background pixels. In doing so, we obtained a binary mask on which we applied morphological operations (dilation followed by erosion) to close the small holes in the foreground. Finally, we took the connected components with areas above a certain threshold and fit the minimum rectangular bounding boxes around them. An example is shown in Figure 2(f).

### Implementation Details

1. **Experiments Setup:** We trained the proposed model for both a three-way classification (i.e., $K = 3$ for NP, CAP and COVID-19) and three binary classification tasks ($K = 2$), i.e., NP vs. COVID-19, NP vs. CAP and CAP vs. COVID19, respectively. In the three-way classification settings, we first trained individual classifiers at different convolution blocks. In our experiment, we chose Conv3, Conv4 and Conv5, respectively. Then, we trained a joint classifier on the aggregated prediction scores (as described in the "Multi-Scale Learning" Section). All the classifiers were trained with the loss in Eq. 1. Finally, we conducted a 5-fold cross-validation on all tasks that in each category, we split the datasets into training, validation and test set. This can ensure that no samples (images) originating from validation and test patients were used for training. In each fold, we held out ~20% of all samples for validation and test, and the remaining were used for training.
2. **Training Configurations:** We implemented the proposed model (as depicted in Figure 1) using Tensorflow 1.14.0. All models were trained from scratch on four Nividia GeForce GTX 1080 Ti

GPUs with an Adam optimiser (learning rate: $10^{-4}$, $\beta_1 = 0.5$, $\beta_2 = 0.9$ and $\epsilon = 10^{-8}$). We set $\gamma$ to 1 in the focal modulator $f$ and the total number of training iterations was set to 20,000. Early stopping was enabled to terminate training automatically when validation loss stopped decreasing for 1,000 iterations. We run validation once every 500 iterations of training, a checkpoint was saved automatically if the current validation accuracy exceeded the previous best validation accuracy. Once the training was terminated, we generated a frozen graph on the latest checkpoint and saved it in .pb format. For testing, we simply loaded the frozen graphs and retrieved the required nodes. Empirically, we found that 20 to 30 steps were good enough to approximate the integral when computing the integrated gradients; thus, we fix $m = 25$ in Eq. 3.

3. **Data Augmentation:** We applied several random on-the-fly data augmentation strategies during training, including (1) cropping square patches at the centre of the input frames with a scaling factor randomly chosen between 0.7 to 1, and resized the crops to the size of 224×224 (input resolution); (2) rotation with an angle randomly selected within $\theta = -25^o$ to $25^o$; (3) Random horizontal reflection, i.e., flipped the images in the left-right direction with a probability $p$ = 0.5; and (4) adjust contrast by randomly darkening or brightening with a factor ranging between 0.5 and 1.5.

### Evaluation Metrics

Using positive results of the RTPCR testing as the ground truth labelling for the COVID-19 group and diagnosis results of CAP and NP patients, accuracy, precision, sensitivity and specificity [32,33] of our classification framework were calculated. We also carried out the area under the receiver operating characteristic curve (AUC) analysis for the quantification of our classification performance. For the lung segmentation, we used Dice score [34] to evaluate the accuracy.

---

## EXPERIMENTS AND RESULTS

### Lung Segmentation

In order to evaluate the lung segmentation network, we randomly split the 60 TCIA data with ground truth into 40 training, 10 validation and 10 independent testing datasets. Ablation study results of different pre-processing and post-processing methods using Dice scores are shown in Figure 3.

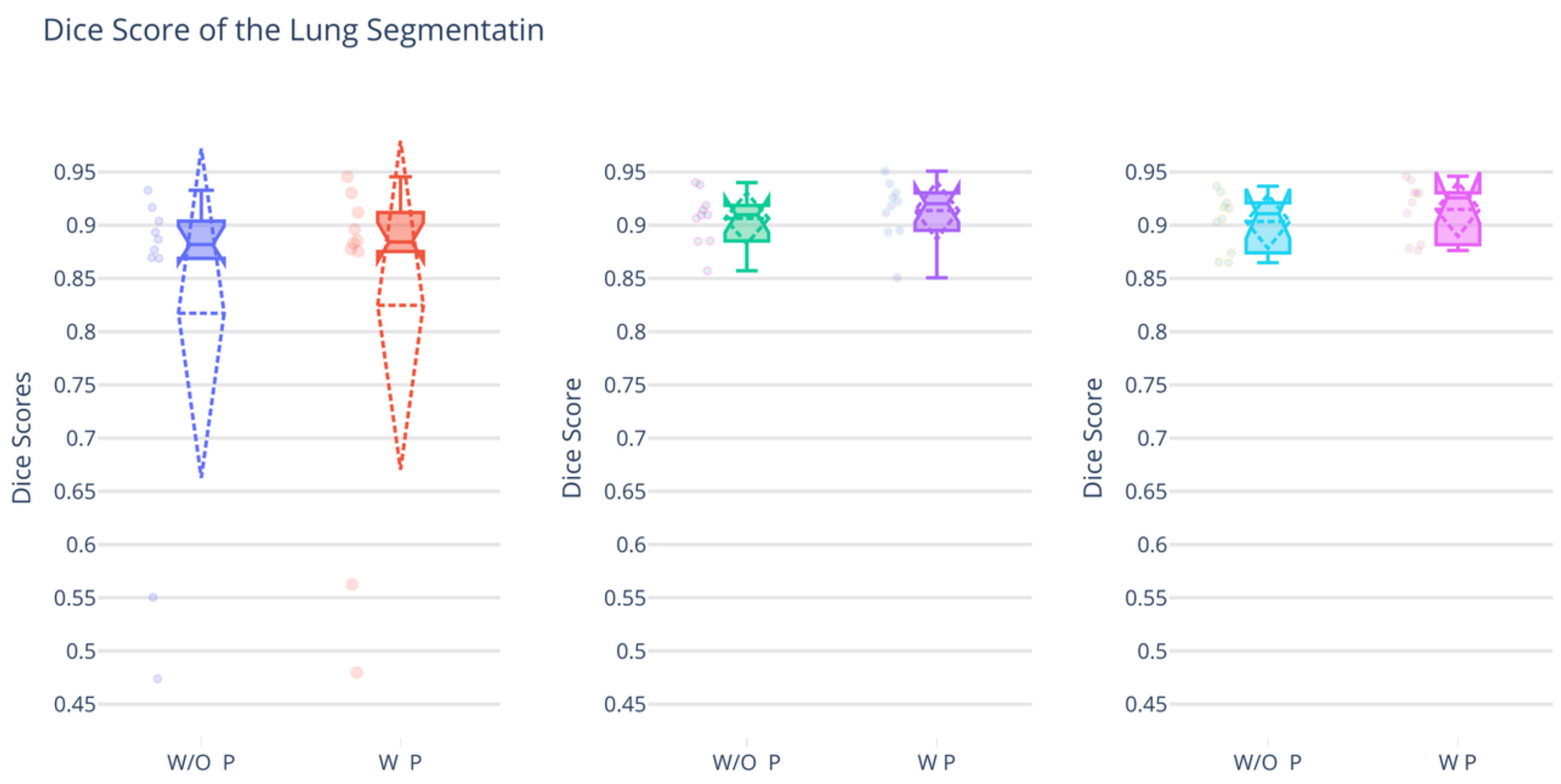

*Figure 3.: Dice scores of the lung segmentation using different pre-processing and post-processing methods on the TCIA dataset. Left Panel: without any pre-processing; Middle Panel: normalising using a pre-defined Hounsfield unit (HU) window; Right Panel: normalising using the proposed fixed-sized sliding window. W/O P: without multi-view learning based post-processing; W P: with multi-view learning based post-processing.*

**Infection Detection**

**A. Class Activation Mapping**

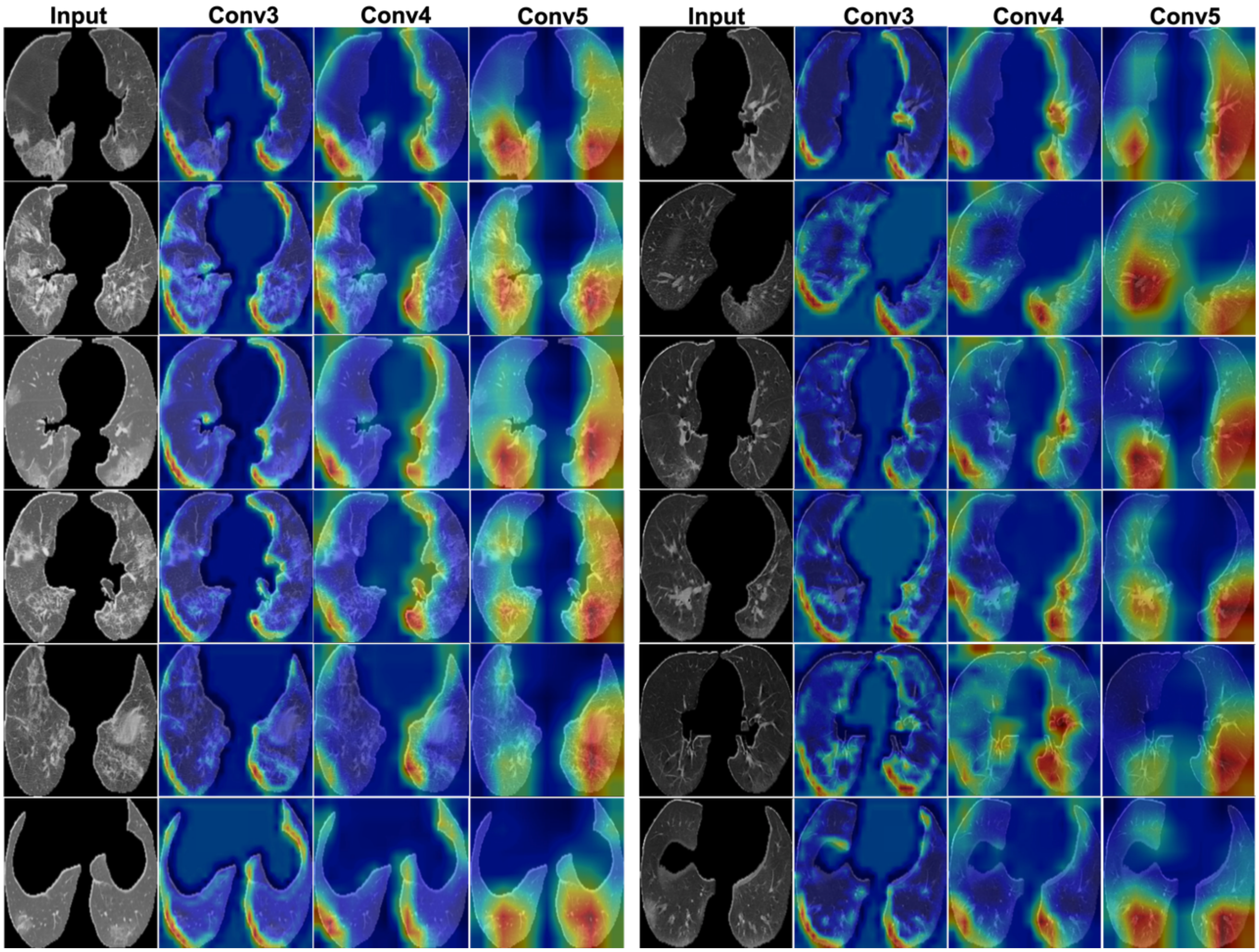

*Figure 4.: Results of the multi-scale COVID-19 class activation mapping. COVID-19: coronavirus disease 2019.*

As a result of multi-scale learning, Figure 4 illustrates some examples of COVID-19 class activation maps (CAMs) obtained at the different feature levels, i.e., Conv3, Conv4 and Conv5. The hot areas indicate where infections happen. The hotter the areas, the more likely they are infected. Of note from the multi-scale CAMs, our proposed model learns to capture the distributions of lesions with different scale: for instance, the large patchy-like lesions, such as crazy paving sign and consolidation; and also small nodule-like lesions, such as ground-glass opacities (GGO) and bronchovascular thickening. Notably, we found the mid-level layers, i.e., Conv3 and Conv4, learn to detect small lesions (GGO most frequently), especially those distributed peripherally and subpleurally. However, they are not able to capture larger patchy-like lesions, and this may be because of the limited receptive field at the mid-layers. In contrast, the high-level layer, i.e., Conv5, having sufficiently large receptive filed learns well to detect the large patchy-like lesions, such as crazy paving sign and consolidation, which are often distributed centrally and peribronchially.

**B. Categorical-Specific Saliency**

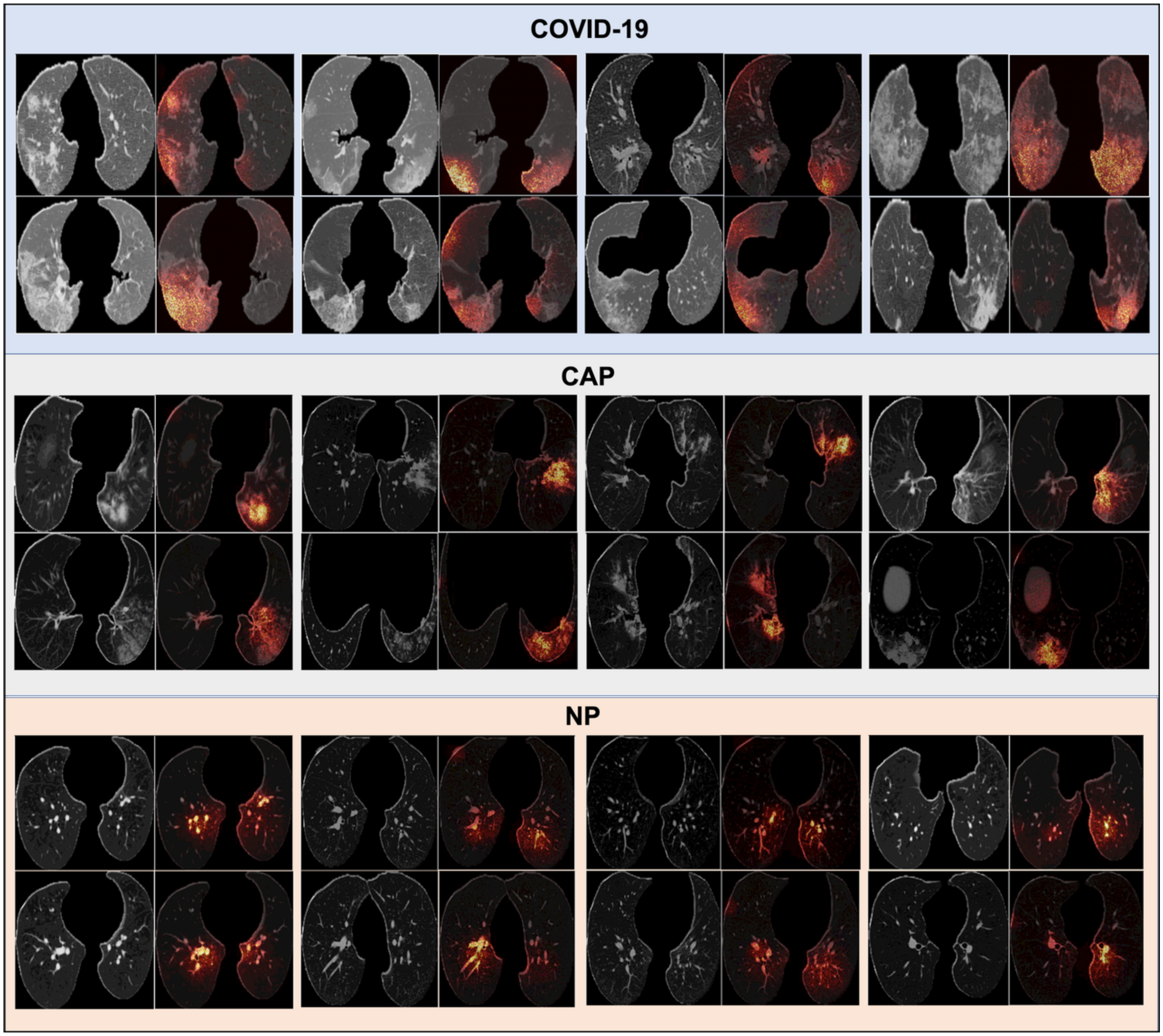

*Figure 5.: Results of the categorical-specific joint saliency. COVID-19: coronavirus disease 2019, CAP: Community Acquired Pneumonia, NP: Non-Pneumonia.*

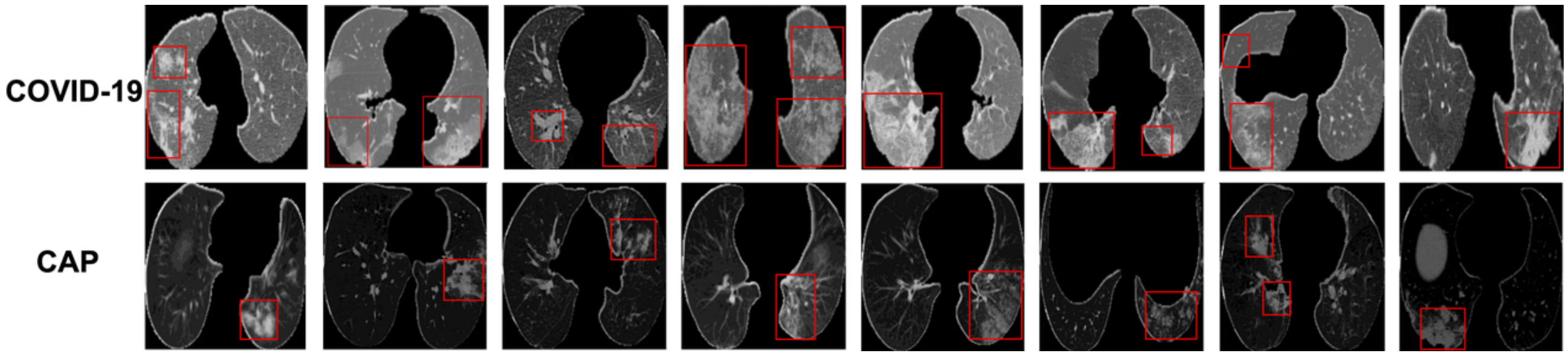

*Figure 6.: Bounding boxes extracted from saliency for COVID-19 and CAP examples. (Corresponding to the examples in Figure. 5). COVID-19: coronavirus disease 2019, CAP: Community Acquired Pneumonia.*

Figure 5 shows the examples of categorical-specific joint saliency computed by integrated gradients. It shows the original inputs on the left and the overlaid saliency on the right. CAMs showed in Figure 4 only depict the spatial distribution of infection. However, it can not be used for precise localisation of the lesions. The saliency maps, on the other hand, can provide pixel-level information that delineates the exact extent of the lesions so providing a precise localisation of the lesions.

Furthermore, clinically, this can also be useful for diagnosis that with the saliency maps, we can estimate the percentage of infection to lung areas. These saliency maps highlight the pixels that contribute to increasing categorical-specific scores: the brighter the pixels, the more significant the contribution. Intuitively, one can also interpret this as the brighter the pixels are, the more critical features to the network to make the decision (prediction). It is of note that in Figure 4 and Figure 5, there is not only an inter-class contrast variation (due to the data are collected from multi-institutions) but also an intra-class contrast variation, especially in COVID-19 group. In our experiments, we found that histogram matching can suppress lesions, especially on COVID-19 images; for instance, GGO disappears or become less apparent. Besides, this leads to inferior performance of detection. Therefore, instead of directly applying histogram matching, we applied random on-the-fly contrast adjustment for data augmentation at training time. This turns out to be very effective, as demonstrated in Figure 5, our proposed model learns to be invariant to image contrast, and precisely capture the lesions.

In addition, from the COVID-19 and CAP saliency, we found that the CAP lesions are generally smaller and more constrained locally compare to COVID-19 cases that often have multiple infected regions and lesions are massive and scattered. It should also be noted that COVID-19 and CAP lesions do share similar radiographic features, such as GGO and air space consolidation. Besides, GGOs appear frequently in subpleural regions as well in CAP cases. Interestingly, from the saliency map for the NP cases, we found the network takes the pulmonary arteries as the salient feature. Finally, Figure 6 shows the bounding boxes extracted from COVID-19 and CAP saliency maps (corresponding to the examples in Figure 5). We found the results agree with our primary findings that CAP cases have less infected areas and often there is single-instance of infection, in contrast, COVID-19 cases often have more infected areas (multi-instances of infection), and the COVID-19 lesions vary a lot in terms of extent. Overall, CAP infection areas are smaller compare to those of COVID-19.

### Classification Performance

Table 2: The overall classification performance comparison between different tasks on the test set.

| Tasks | Accuracy (%) | Precision (%) | Sensitivity (%) | Specificity (%) | AUC |
|---|---|---|---|---|---|
| Conv3 auxiliary classifier*† | 72.3 [69.9, 75.1] | 73.3 [70.9, 75.7] | 71.2 [68.8, 73.8] | 70.3 [68.0, 72.6] | 0.742 [0.720, 0.764] |
| Conv4 auxiliary | 83.2 [80.9, | 83.9 [81.3, | 82.5 [79.8, | 81.7 [79.4, | 0.834 |
| classifier*† | 85.5] | 86.5] | 85.2] | 84.0] | [0.813, 0.855] |
| Conv5 auxiliary classifier*† | 84.3 [82.4, 86.2] | 84.1 [82.0, 86.2] | 83.8 [81.4, 86.2] | 81.3 [78.8, 82.8] | 0.835 [0.809, 0.861] |
| Joint Classifier*‡ | 87.4 [84.4, 90.3] | 87.5 [85.1, 89.8] | 88.5 [85.9, 91.0] | 87.1 [84.6, 89.6] | 0.895 [0.866, 0.926] |
| NP/COVID-19≀‡ | 96.2 [94.5, 97.9] | 97.3 [95.9, 98.7] | 94.5 [93.2, 95.8] | 95.3 [93.9, 96.7] | 0.970 [0.957, 0.983] |
| NP/CAP≀‡ | 94.0 [92.5, 95.5] | 95.1 [93.5, 96.7] | 92.0 [90.3, 93.7] | 92.5 [90.9, 94.1] | 0.952 [0.935, 0.969] |
| COVID-19/CAP≀‡ | 89.1 [87.2, 91.0] | 91.5 [90.1, 93.0] | 87.0 [85.9, 88.1] | 86.2 [84.6, 87.8] | 0.906 [0.886, 0.926] |
| NP/COVID-19 (NTS-NET)≀‡ | 90.6 [94.4, 87.8] | 74.1 [75.7, 70.2] | 83.3 [86.8, 81.7] | 95.6 [96.9, 91.7] | 0.943 [0.964, 0.907] |
| NP/CAP (NTS-NET) ≀‡ | 85.8 [89.1,83.3] | 79.5 [83.0, 77.8] | 77.8 [80.4, 74.1] | 88.9 [90.3, 86.1] | 0.911 [0.944, 0.957] |
| COVID-19/CAP (NTS-NET)≀‡ | 84.9 [86.1,81.5] | 77.8 [81.0, 75.3] | 81.4 [82.4, 79.7] | 89.2 [92.1, 86.7] | 0.864 [0.889, 0.924] |

* Note: values in brackets are 95% confidence intervals [95%CI,%]. AUC: area under the receiver operating characteristic curve, COVID-19: coronavirus disease 2019, CAP: Community Acquired Pneumonia, NP: Non-Pneumonia. *: three-way classification tasks (i.e., NP/CAP/COVID-19). ≀: binary classification tasks.† : single-scale learning. ‡: multi-scale learning.

Performance of our proposed model for each specific task was evaluated with 5-fold cross-validation, and the results on the test set are reported and summarised in Table 2. We use five evaluation metrics, which are accuracy (ACC), precision (PRC), sensitivity (SEN), specificity (SPE) and the area under the ROC curve (AUC). We report the mean of 5-fold cross-validation results in each metric with the 95% confidence interval. We also compared our proposed method with a reimplementation of the Navigator-Teacher-Scrutinizer Network (NTS-NET) [35].

As described earlier in the experimental settings, basically we have two groups of tasks: three-way classification tasks (indicated by *) and binary classification tasks (indicated by ≀), and two learning configurations: single-scale learning (indicated by †) that assigns an auxiliary classifier to a specific feature level, and multi-scale learning (indicated by ‡) that aggregates the multi-level prediction scores then trained with a joint classifier. All the binary tasks listed were trained with the multi-scale learning. In terms of three-way classification, we found the multi-scale learning with joint classifier achieves superior overall performance than any of the single-scale learning tasks. It is of note that among the single-scale learning tasks, classification with Conv4 and Conv5 features achieve very similar performance in every metric, which is significantly better than classification with mid-level, i.e., Conv3 features. One possible explanation is the mid-level features are not sufficiently semantic compare to higher-level features, i.e., Conv4 and Conv5. As we know, high-level CNN representations are semantically strong but poorly at preserving spatial details, whereas mid-lower level CNN representations preserve well the local features but lack of semantic information.

Furthermore, it is of note that, overall, binary classification tasks achieve significantly better performance than three-way classification, especially in the tasks, such as NP/COVID-19 and NP/CAP. It can be seen our proposed model is reasonably good at distinguishing COVID-19 cases from NP cases as suggested by the results, showing that it achieves a mean ACC of 96.2%, PRC of 97.3%, SEN of 94.5%, SPE of 95.3% and AUC of 0.970, respectively. One can explain this is because binary classification is less complicated, and there is also less uncertainty than three-way classification. This may also because COVID-19 and CAP image features are intrinsically discriminative compare to the NP cases. For instance, as the COVID-19 cases demonstrated earlier, there is often a combination of various diseased patterns and large areas of infection on the scans.

Last but not least, we found that the performance of COVID-19/CAP classification is the least superior among all the binary classification tasks. One possible reason is COVID-19 shares the similar radiographic features with CAP, such as GGO and airspace consolidation and the network capacity may not be enough to learn disease-specific representations. Nevertheless, the results obtained using our proposed method outperformed the ones obtained by the NTS-NET.

Table 3: The performance (breakdown into each individual class) of three-way classification on the test set.

| Categories | Accuracy (%) | Precision (%) | Sensitivity (%) | Specificity (%) | AUC |
|---|---|---|---|---|---|
| COVID-19 | 89.2 [87.1, 91.3] | 87.9 [85.8, 90.0] | 88.6 [86.5, 90.7] | 87.6 [85.0, 90.2] | 0.923 [0.897, 0.949] |
| CAP | 84.7 [80.6, 88.8] | 82.3 [78.9, 85.0] | 87.5 [83.8, 91.2] | 83.0 [79.8, 86.2] | 0.864 [0.832, 0.896] |
| NP | 88.3 [85.7,90.9] | 92.5 [90.6,94.4] | 89.5 [87.6, 91.1] | 91.3 [89.5, 93.1] | 0.901 [0.870, 0.932] |
| COVID-19 (NTS-NET) | 84.3[86.2,82.9] | 72.4[73.6,70.3] | 76.4[77.4,73.1] | 89.8[92.1,88.4] | 0.912[0.951,0.899] |
| CAP | 83.2[85.7,81.1] | 70.7[72.2,68.4] | 74.5[78.2,70.8] | 89[91.7,85.8] | 0.884[0.909,0.857] |
| (NTS-NET) | | | | | |
| NP (NTS-NET) | 80.1[81.2,77.7] | 66.7[68.2,62.7] | 73.8[75.9,72.1] | 89.6[90.8,87.9] | 0.841[0.854,0.807] |

* Note: values in brackets are 95% confidence intervals [95%CI,%]. AUC: area under the receiver operating characteristic curve, COVID-19: coronavirus disease 2019, CAP: Community Acquired Pneumonia, NP: Non-Pneumonia.

We also break down the overall performance, i.e., the joint classifier (indicated by *‡) into classes, and the classification metrics are reported for each class, as shown in Table 3 and Figure 7. We found that the "COVID-19" and the "NP" classes achieve the comparable performance in each metric and the "NP" class has higher sensitivity (91.3%) than the COVID-19 (87.6%) and CAP (83.0%). Besides, we found, overall, the "COVID" remains the best performed and the most discriminative class with a mean AUC of 0.923, compared to the "CAP" (0.864) and the "NP" (0.901). It can also be noted that the overall results for the class "CAP" are moderately lower than those of the "NP" and "COVID-19". This could be correlated with our finding in the COVID-19/CAP classification that because of similar appearance, the "CAP" class is likely to be misclassified as the "COVID-19" sometimes. Also, another possible reason is that the network could have learned and be distracted by the few "NP noises", and there might be a fractional number of non-infected slices in between the CAP training samples. This is because we sampled all the available slices from each subject, and there might be a few slices having no infections.

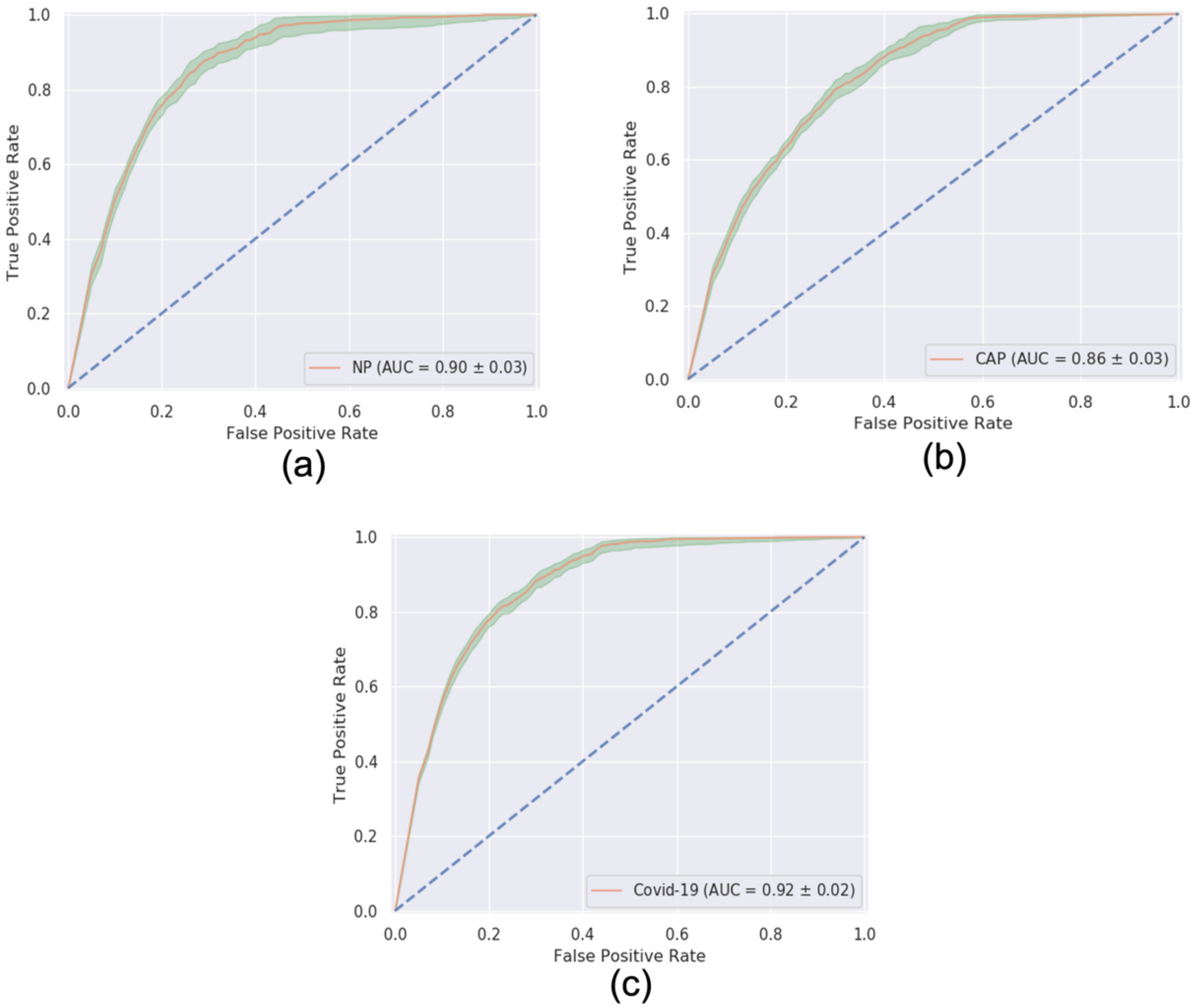

*Figure 7. Receiver operating characteristic (ROC) of individual categories for three-way classification (5-fold cross-validated). (a) NP with AUC of 0.90±0.03 (mean±standard deviation); (b) CAP with AUC of 0.86±0.03 (c) COVID-19 with AUC of 0.92±0.02. The green region indicates the 95%CI. COVID-19: coronavirus disease 2019, CAP: Community Acquired Pneumonia, NP: Non-Pneumonia, CI: Confidence Interval.*

---

## DISCUSSIONS

In this work, we have presented a novel weakly supervised deep learning framework that is capable of learning to detect and localise lesions on COVID-19 and CAP CT scans from image-level label only. Different from other works, we leverage the representation learning on multiple feature levels and have explained what features can be learned at each level. For instance, the high-level representation, i.e., Conv5 captures the patch-like lesions that generally have a large extent. However, it tends to discard small local lesions. This is well complemented by the mid-level representations (Figure 4), i.e., Conv4 and Conv5, from which the lesions detected also correspond to our clinical findings that the infections usually located in the peripheral lung (95%), mainly in the inferior lobe of the lungs (65%), especially in the posterior segment (51%). We speculate that it is mainly because there are more well-developed bronchioles, alveoli, rich blood flows and immune cells such as lymphatic cells in the periphery. These immune cells played a vital role in the inflammation caused by the virus. We have also demonstrated that combing multi-scale saliency maps, generated by integrated gradients, is the key to achieve a precise localisation of multi-instance lesions.

Furthermore, from a clinical perspective, the joint saliency is useful that it provides a reasonable estimation of the percentage of infected lung areas, which is a crucial factor that clinicians take account for evaluating the severity of a COVID-19 patient. Besides, the classification performance of the proposed network has been studied extensively that we have not only conducted three-way classification but also binary classification by combining any two of the classes.

We found one limitation of the proposed network is that it is not discriminative enough when it comes to separate the CAP from COVID-19. We suspect this is due to the limited capacity of the backbone CNN that a straightforward way of boosting CNN capacity is to increase the number of feature channels at each level. Another attempt in the future would be employing more advanced backbone architecture, such as Resnet and Inception. Another limitation in this work is that we have trained the networks on individual slices (images) that we use all available samples for each subject. However, for the CAP or COVID-19 subjects, there might be fractional non-infection slices in between which could introduce noises in training. In the future, we can address the limitation by attention-based multiple instances learning that instead of training on individual slices, we put the patient-specific slices into a bag and train on bags. The network will learn to assign weights to individual slices in a COVDI-19 or CAP positive bag and automatically sample those high weighted slices for infection detection.

---

## CONCLUSION

In this study, we designed a weakly supervised deep learning framework for fast and fully-automated detection and classification of COVID-19 infection using retrospectively extracted CT images from multi-scanners and multi-centres. Our framework can distinguish COVID-19 cases accurately from CAP and NP patients. It can also pinpoint the exact position of the lesions or inflammations caused by the COVID-19, and therefore can also potentially provide advice on patient severity in order to guide the following triage and treatment. Experimental findings have indicated that the proposed model achieves high accuracy, precision and AUC for the classification, as well as promising qualitative visualisation for the lesion detections. Based on these findings we can envisage a large-scale deployment of the developed framework.